\documentclass[letterpaper, 10 pt, conference]{ieeeconf}

\IEEEoverridecommandlockouts

\usepackage{amsmath,amssymb,amsfonts}
\usepackage{graphicx}
\usepackage{booktabs}
\usepackage{multirow}
\usepackage{url}
\usepackage[hidelinks]{hyperref}
\usepackage{cite}
\usepackage{balance}
\usepackage{xspace}

\newcommand{\COtwo}{CO$_2$}

\title{A Multi-Criterion Approach to Smart EV Charging\\with CO$_2$ Emissions and Cost Minimization}

\author{%
Giuseppe C. Calafiore, Luca Ambrosino, Khai Manh Nguyen,\\ Minh Binh Vu, Riadh Zorgati, Laurent El Ghaoui%
\thanks{L. Ambrosino and G. C. Calafiore are with the Department of Electronics and Telecommunications, Politecnico di Torino, Italy. K. M. Nguyen, M. B. Vu and L. El Ghaoui are with VinUniversity, Hanoi, Vietnam. R. Zorgati is with EDF Lab Paris-Saclay, France.}%
}

\begin{document}
\maketitle

\begin{abstract}
We study carbon-aware smart charging in a fossil-dominated grid by coupling a simplified hydro-thermal-renewable dispatch model with a tractable linear charging scheduler. 
The case study is informed by Vietnam's regional data. 
Thermal units remain dominant, renewables are time-varying, and hydropower is modeled through a single reservoir budget. From the day-ahead dispatch we derive hourly carbon intensity and a corresponding carbon-cost signal; these are combined with a local time-of-use tariff in the EV charging problem. The resulting weighted-sum linear program is  multi-objective: by sweeping the trade-off coefficient, we recover the supported Pareto frontier between electricity cost and charging-associated emissions. In a 300-EV public-charging scenario with a 0.8~MW feeder cap, the proposed carbon-aware scheduler preserves the 19.8\% bill reduction of a cost-only optimizer while lowering charging-associated emissions by 7.3\%; a more carbon-focused tuning still remains 12.6\% cheaper and 9.3\% cleaner than a FIFO baseline. A hydro-sensitivity study shows that changing the reservoir budget by $\pm $20\% moves the mean grid carbon intensity from 360 to 466~g/kWh, yet the carbon-aware scheduler remains consistently cheaper and cleaner than FIFO. The dispatch and charging LPs solve in few milliseconds on a standard desktop computer, showing that the framework is lightweight enough for repeated day-ahead studies.
\end{abstract}

\section{Introduction}
Electric vehicles can strongly reduce transport emissions, but the grid-side benefit depends on \emph{when} charging takes place and on the carbon content of the electricity actually used \cite{IEA2024}. This is particularly relevant in systems where coal and gas still provide a large share of daily generation. In such contexts, a charging strategy that reacts only to electricity prices may shift demand toward hours that are cheap for the operator but not necessarily clean for the environment.

The two research threads behind this paper are well established. On the supply side, unit-commitment and dispatch formulations with emission-aware objectives are standard tools for studying cost--carbon trade-offs in hydro-thermal and renewable-rich systems \cite{Zhang2016,Norouzi2014}. On the demand side, smart EV charging is often posed as a linear or convex scheduling problem that reallocates flexible charging load across time \cite{Liu2021,Sun2020,Calafiore2025}. A parallel literature shows that the environmental impact of charging depends strongly on marginal or time-varying supply conditions: Ontario and Great Britain studies quantify large timing effects \cite{Gai2019,Tang2021}, while optimization-based formulations schedule EV demand directly against marginal-emissions or real-time carbon-intensity signals \cite{Tu2020,Cheng2022}. The gap we address is practical rather than theoretical: public charging operators often need a lightweight workflow that turns system-level generation information into a carbon-aware charging signal without solving one large, tightly coupled market-clearing problem.

We therefore adopt a three-step pipeline. First, a simplified hourly dispatch model computes a least-cost Vietnam-inspired hydro-thermal-renewable mix over one representative day. Second, the dispatch is converted into hourly carbon intensity and carbon cost. Third, these quantities are combined with a local time-of-use (TOU) tariff in a linear charging scheduler. 
The contribution is not a new class of unit-commitment models; rather, it is a compact and reproducible way to merge (i) thermal, renewable and hydro resources, (ii) a locally consistent tariff instead of foreign wholesale prices, (iii) an optimization-based cost-only baseline besides FIFO, (iv) an explicit Pareto analysis, and (v) computational-burden reporting.

The decoupled architecture is appropriate when the charging station behaves as a \emph{price taker}. In the experiments, the station peak is 0.8~MW whereas the representative system peak is 35.5~GW, hence the charging hub accounts for only $2.25\times 10^{-5}$ of system peak demand. For larger aggregations, the coupling would need to be iterated or formulated as a bilevel/fixed-point problem; we return to this in the discussion.

\section{Vietnam-Inspired Dispatch Layer}
\subsection{Generation mix and representative profiles}
The supply-side model integrates thermal sources, renewables, hydropower, imports, and associated \COtwo{} emissions.  Table~\ref{tab:source_data} reports the source shares used to scale installed capacity from official Vietnam Electricity (EVN, the state-owned national electric utility of Vietnam) data, together with dispatch costs and life-cycle emission factors from the sources used in the original study \cite{EVN2024Mix,WNA2024}. Monetary quantities are  handled in $\mathrm{VND/MWh}$.

\begin{table}[t]
\caption{Source data used in the dispatch LP.}
\label{tab:source_data}
\centering
\footnotesize
\begin{tabular}{lccc}
\toprule
Source & Cap. share [\%] & Cost [$10^6$ VND/MWh] & \COtwo{} [g/kWh] \\
\midrule
Coal   & 33.2  & 2.100 & 820 \\
Gas    & 8.9   & 1.428 & 490 \\
Fuel   & 1.4   & 3.000 & 740 \\
PV     & 18.55 & 2.046 & 48 \\
Wind   & 8.25  & 2.086 & 12 \\
Hydro  & 28.4  & 1.128 & 24 \\
Import & 1.2   & 2.200 & 300 \\
\bottomrule
\end{tabular}
\end{table}

The demand trajectory is a representative 24-hour system profile spanning 26.1--35.5~GW. Photovoltaic (PV) availability is modeled through a daylight bell shape, while wind follows a milder diurnal profile consistent with Vietnam-scale wind modeling based on reanalysis-driven virtual wind farms \cite{Staffell2016}. Hydropower is aggregated into a single energy-equivalent reservoir, following the usual virtual-reservoir simplification employed in system studies and motivated by the importance of hydro flexibility in Vietnam \cite{NguyenTien2018}. This aggregation keeps the hydro discussion in the paper while avoiding plant-level binaries that are not needed for the charging application.

Thermal units (coal, gas, fuel, oil) are assumed fully available up to their aggregate capacity, so the thermal layer plays the role of the dispatchable backbone. Fuel oil therefore appears as a costly emergency option rather than a routinely used source. Renewable production is curtailable and no stationary battery is included, which means that midday PV surplus cannot be shifted to the evening peak. The hydro parameters are set to $s_0=110$~GWh, $\underline s=30$~GWh, $\overline s=160$~GWh, with a flat inflow of 7~GWh/h; these values keep hydro energy in the same order of magnitude as its annual contribution while still leaving enough freedom to arbitrage between midday and evening demand.

\begin{figure}[t]
    \centering
    \includegraphics[width=0.96\columnwidth]{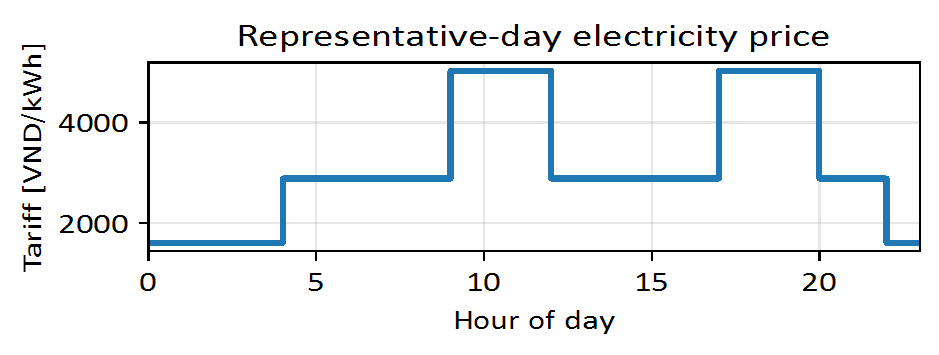}
    \caption{Representative-day retail electricity price, based on the EVN business-customer time-of-use tariff.}
    \label{fig:price}
\end{figure}

The representative-day retail electricity price profile in Figure~\ref{fig:price} is constructed from the Vietnam Electricity business-customer retail tariff under Decision No.~1279/QD-BCT dated 9 May 2025, using the published off-peak, standard, and peak rates together with EVN's time-of-use period definitions; the official half-hour boundaries are mapped to our hourly discretization, \cite{EVNTariff2025,EVNTOU2016}.
For the weekday representative day considered here, off-peak hours are taken overnight, peak hours occur in the late morning and early evening, and the remaining hours are standard-price periods, consistent with EVN's published TOU schedule \cite{EVNTOU2016}.

\begin{figure}[t]
    \centering
    \includegraphics[width=0.98\columnwidth]{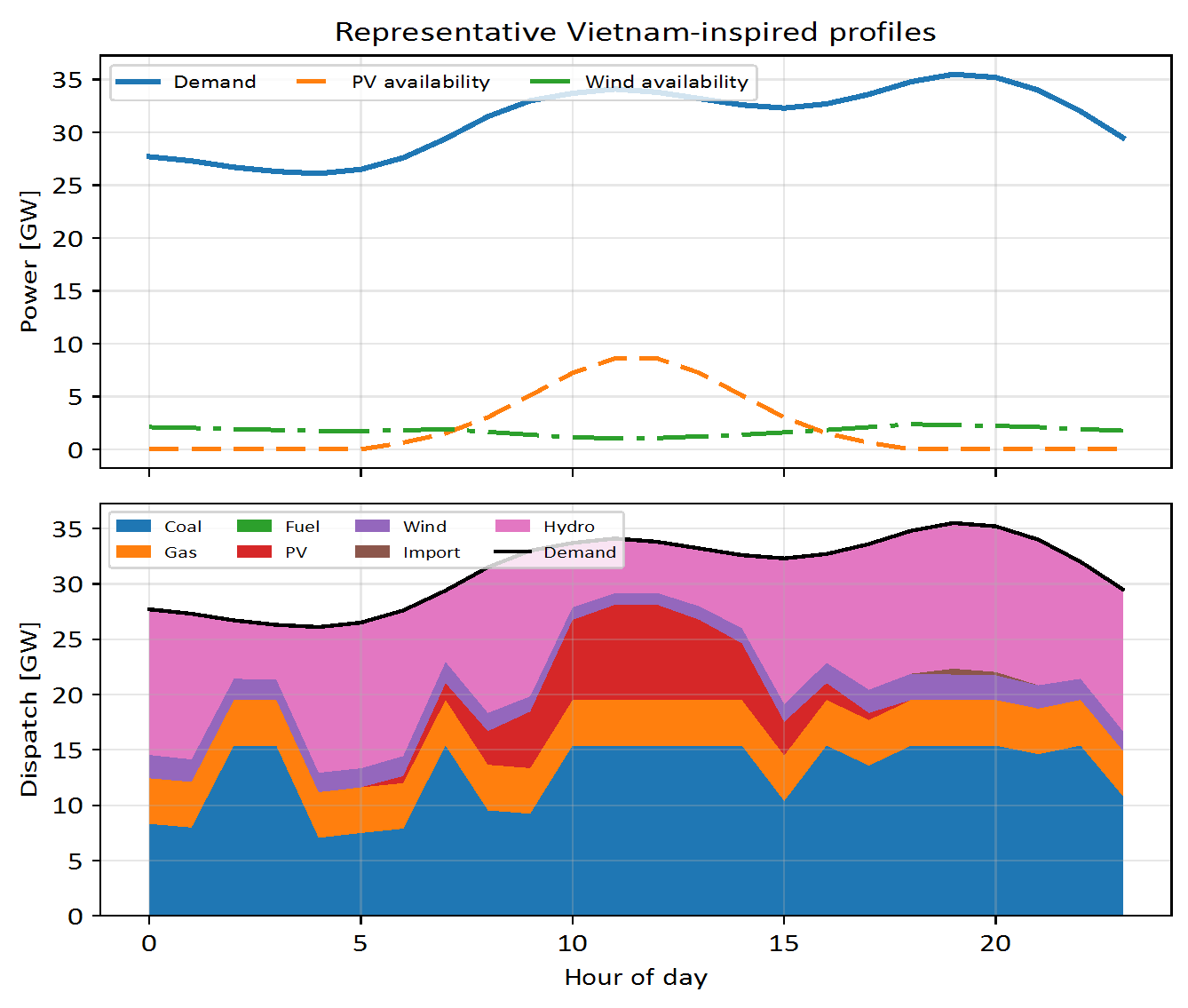}
    \caption{Representative demand and renewable availability (top), and optimal hourly dispatch mix (bottom).}
    \label{fig:dispatch}
\end{figure}

\subsection{Hydro-thermal-renewable dispatch model}
Let $p_{t,j}$ denote the power produced at hour $t\in\{1,\ldots,24\}$ by source $j\in\mathcal{G}:=\{$coal, gas, fuel, PV, wind, import, hydro$\}$. For thermal units and imports, $0\le p_{t,j}\le \bar p_{t,j}$ with time-invariant upper bounds. For PV and wind, $\bar p_{t,j}$ is hour-dependent. Hydropower output is denoted $h_t:=p_{t,\mathrm{hydro}}$, and the reservoir state $s_t$ is measured in energy-equivalent MWh. 
Problem \eqref{eq:dispatch} below is the day-ahead least-cost dispatch model: over the representative 24-hour horizon, it determines the hourly generation mix that satisfies demand while respecting source-capacity constraints and hydro-storage dynamics.
\begin{equation}
\begin{aligned}
\min_{p,s}\;& \sum_{t=1}^{24}\sum_{j\in\mathcal{G}} c_j p_{t,j} \\
\text{s.t. }& \sum_{j\in\mathcal{G}} p_{t,j} = D_t, \quad t=1,\ldots,24,\\
& s_t = s_{t-1}+\phi_t-h_t, \quad t=1,\ldots,24,\\
& \underline s\le s_t\le \overline s,\quad 0\le p_{t,j}\le \overline p_{t,j}.
\end{aligned}
\label{eq:dispatch}
\end{equation}
Here, $p:=\{p_{t,j}\}_{t=1,\ldots,24;\,j\in\mathcal G}$ and $s:=\{s_t\}_{t=1}^{24}$ denote the decision variables, where $p_{t,j}$ is the generation scheduled from source $j$ at hour $t$, and $s_t$ is the hydro-reservoir energy level at hour $t$. The parameter $c_j$ denotes the unit generation cost of source $j$ (in VND/MWh), $D_t$ is the system demand at hour $t$, and $\overline p_{t,j}$ is the available capacity of source $j$ at hour $t$ (constant for thermal units and imports, time-varying for PV and wind). Moreover, $\phi_t$ is the exogenous hydro inflow during hour $t$, while $\underline s$ and $\overline s$ are the minimum and maximum admissible reservoir levels, respectively; $s_0$ is the given initial reservoir level. Finally, since hydropower is included in $\mathcal G$, one has $h_t=p_{t,\mathrm{hydro}}$.
With a one-hour time discretization, $p_{t,j}$, $h_t$, and $\phi_t$ can be interpreted equivalently as hourly average powers or as hourly energies.

All costs are linear because we seek a lightweight proxy for carbon-aware charging, not a full mixed-integer unit-commitment problem (UCP)  with startup logic. 
This simplification is intentional: the charging layer only needs an hourly signal indicating the likely marginal carbon intensity of electricity supply, not the exact commitment of every individual plant. Accordingly, \eqref{eq:dispatch} should be interpreted as a dispatch-based surrogate for carbon-intensity estimation rather than as a full market-clearing simulator.

A second modeling point concerns emissions interpretation. Since the charging station is small at the scale considered here, the carbon signal derived from \eqref{eq:dispatch} is used as an \emph{associated} emission factor for each hour, not as a strict marginal-emission oracle. The charging schedule reshapes the station load, but at this scale it does not materially change the national mix itself. Figure~\ref{fig:dispatch} shows the resulting schedule. Hydropower is heavily used around the evening peak, PV depresses midday coal usage, and fuel oil is essentially absent because of its high cost. Over the representative day, the model dispatches 41.0\% coal, 13.2\% gas, 33.1\% hydro, 7.0\% PV, 5.6\% wind, and negligible fuel/imported power.

\section{From Dispatch to Carbon-Aware Charging}
\subsection{Carbon signal}
From the dispatch solution we compute the hourly grid carbon intensity
\begin{equation}
\kappa_t=\frac{\sum_{j\in\mathcal{G}} \gamma_j p_{t,j}}{\sum_{j\in\mathcal{G}} p_{t,j}},\qquad t=1,\ldots,24,
\label{eq:kappa}
\end{equation}
where $\gamma_j$ is the source-specific emission factor from Table~\ref{tab:source_data}. In the representative day, $\kappa_t$ ranges from 308 to 562~g\COtwo{}/kWh. To turn it into a monetary penalty, we use a reference carbon price $\tau$ and define
\begin{equation}
\pi^{\mathrm{CO}_2}_t = \tau\,\kappa_t/10^6 \qquad [\text{VND/kWh}],
\label{eq:carbon-price}
\end{equation}
which yields an hourly carbon-cost signal. We set $\tau=1.785\times 10^6$~VND/t\COtwo{}, consistent with a 70~USD/t reference level discussed in carbon-pricing studies \cite{IMF2019}.

\subsection{Smart charging LP and Pareto interpretation}
We consider a public station serving $N=300$ EVs over a 24-hour horizon with 10-minute slots. Vehicle $i$ is present on the interval $\mathcal{T}_i$, requests $E_i$~kWh, and can receive at most $\overline p=7.2$~kW per slot. Let $y_{i,t}$ be the energy assigned to EV $i$ in slot $t$, and let $C=0.8$~MW be the feeder cap. The charging problem is
\begin{equation}
\begin{aligned}
\min_{y}\;& \sum_{t=1}^{T} \left(\pi^e_t + \lambda\pi^{\mathrm{CO}_2}_t\right)\sum_{i=1}^{N} y_{i,t} \\
\text{s.t. }& \sum_{t\in\mathcal{T}_i} y_{i,t}=E_i,\quad i=1,\ldots,N,\\
& \sum_{i=1}^{N} y_{i,t}\le C\Delta,\quad t=1,\ldots,T,\\
& 0\le y_{i,t}\le \overline p\,\Delta,\quad t\in\mathcal{T}_i,
\end{aligned}
\label{eq:charge}
\end{equation}
where $\Delta=1/6$~h and $\pi^e_t$ is the TOU retail tariff faced by the station operator. In the experiments, $\pi^e_t$ follows the published EVN business-customer TOU bands, with off-peak hours overnight, standard pricing through most daytime hours, and peak periods in the late morning and early evening \cite{EVNTariff2025,EVNTOU2016}. 

The coefficient $\lambda\ge 0$ is not an ad hoc tweak but the usual weighted-sum scalarization parameter: sweeping it recovers the supported Pareto frontier of the convex bi-objective problem.

A FIFO baseline is used for comparison, but we also compare against the optimization-based cost-only scheduler obtained from \eqref{eq:charge} by setting $\lambda=0$. 
The FIFO baseline is a non-anticipative first-in-first-out charging rule. Specifically, EVs are ordered according to their arrival times, and at each hour the available charging power is assigned first to the earliest-arrived vehicles that are still connected and not yet fully charged. Charging is continued for each such vehicle up to its per-vehicle charging limit or remaining energy requirement, and any residual charging capacity is then passed to the next vehicle in the queue. Hence, FIFO does not optimize against electricity price or carbon intensity; it simply prioritizes vehicles in order of arrival while enforcing the same connection windows, battery requirements, and aggregate charging constraints as in \eqref{eq:charge}.

\section{Numerical Results}
\subsection{Simulation setup}
The charging horizon contains $T=144$ slots of 10 minutes each. We generate $N=300$ EVs through a two-cluster arrival model: half of the vehicles belong to a morning peak and half to an afternoon/evening peak. Parking durations are random, and each energy request is a beta-distributed fraction of the energy that could be delivered within the corresponding availability window. This keeps every instance feasible while producing heterogeneous flexibility levels across users. The station is modeled as a public charging hub with identical AC ports and a single feeder bottleneck.

\begin{table}[t]
\caption{Charging-simulation parameters.}
\label{tab:setup}
\centering
\footnotesize
\begin{tabular}{lc}
\toprule
Quantity & Value \\
\midrule
EVs per day & 300 \\
Time step $\Delta$ & 10 min \\
Per-EV limit $\overline p$ & 7.2 kW \\
Feeder cap $C$ & 0.8 MW \\
TOU tariff [off/std/peak] & 1609 / 2887 / 5025 VND/kWh \\
Reference carbon price $\tau$ & $1.785\times 10^6$ VND/t\COtwo{} \\
Mean delivered energy & 3.82 MWh/day \\
\bottomrule
\end{tabular}
\end{table}

The tariff values in Table~\ref{tab:setup} correspond to the EVN business-customer TOU bands \cite{EVNTariff2025}. 
Our setup keeps the local charging economics grounded in Vietnamese retail conditions, while letting the dispatch layer provide a grid-consistent carbon signal.

\subsection{Representative day charging profiles}
\begin{figure}[t]
    \centering
    \includegraphics[width=0.98\columnwidth]{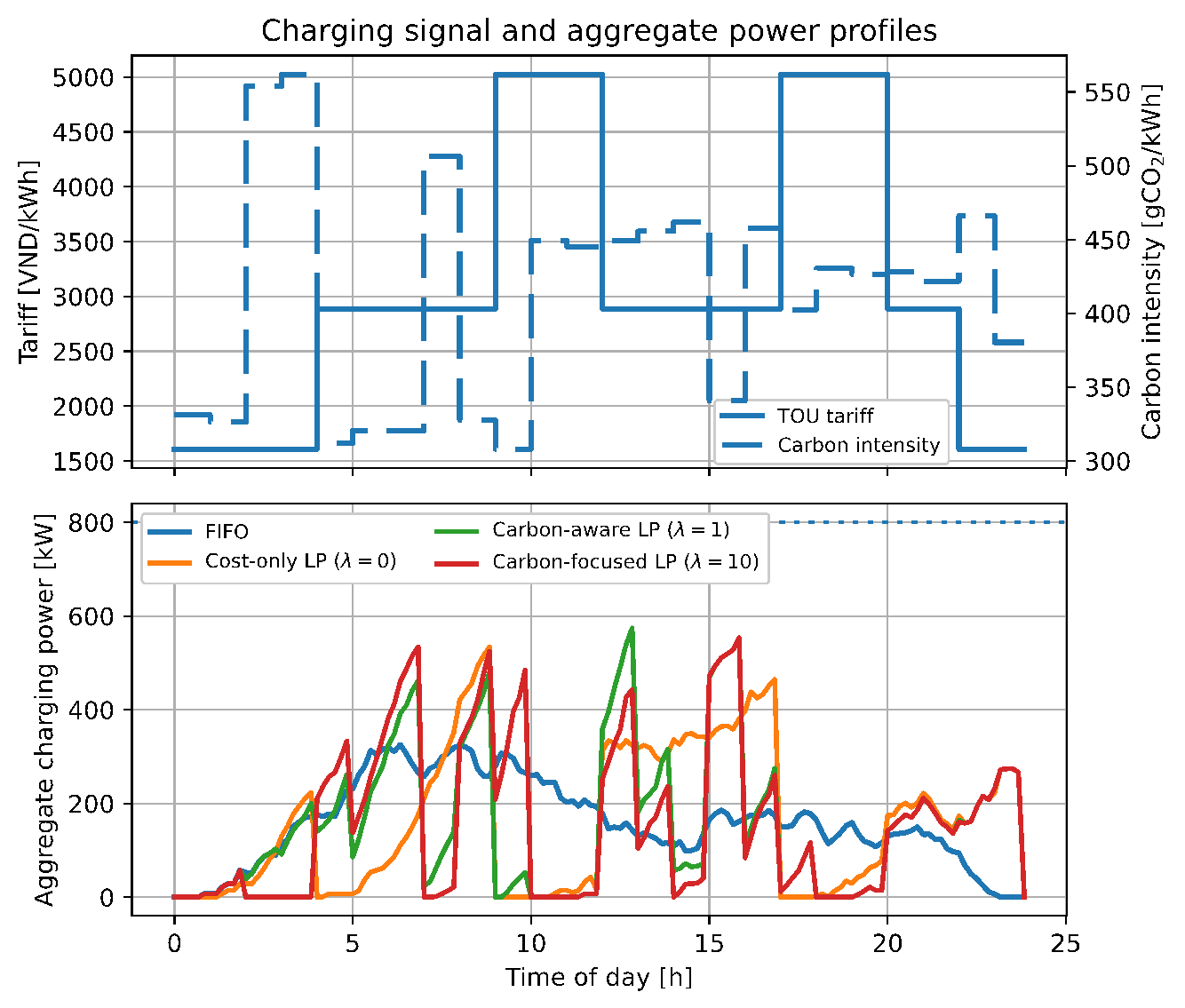}
    \caption{TOU tariff and dispatch-derived carbon signal (top), and aggregate charging power for four policies (bottom).}
    \label{fig:charge}
\end{figure}

In Figure~\ref{fig:charge}, the top panel reports the TOU tariff and the dispatch-derived carbon intensity; the bottom panel shows the aggregate charging power under FIFO, a cost-only LP, a balanced carbon-aware LP ($\lambda=1$), and a more carbon-focused scheduler ($\lambda=10$).

The qualitative behavior can be easily interpreted:  the FIFO policy reacts only to vehicle arrivals and therefore charges whenever demand appears. The cost-only LP shifts energy toward off-peak price bands, especially at night and outside the evening peak. Because the cheapest hours are not always the ``cleanest'' ones, this policy actually \emph{increases} charging-associated emissions. The carbon-aware policies push additional energy toward low-carbon windows around late morning and early afternoon, when coal is partially displaced by PV and hydro.

On the representative day, FIFO yields a 13.09~MVND bill and 1536.6~kg of charging-associated \COtwo{}. The cost-only LP cuts the bill to 10.36~MVND but raises emissions to 1620.4~kg. By contrast, the balanced carbon-aware LP with $\lambda=1$ achieves the \emph{same} bill, 10.36~MVND, while reducing emissions to 1508.8~kg. A more aggressive setting, $\lambda=10$, lowers emissions further to 1404.8~kg with a moderate bill increase to 11.34~MVND. Hence, in this scenario the multi-criterion formulation is useful even relative to another optimizer, not only relative to FIFO.

\subsection{Charging heatmaps}
\begin{figure}[t]
    \centering
    \includegraphics[width=0.98\columnwidth]{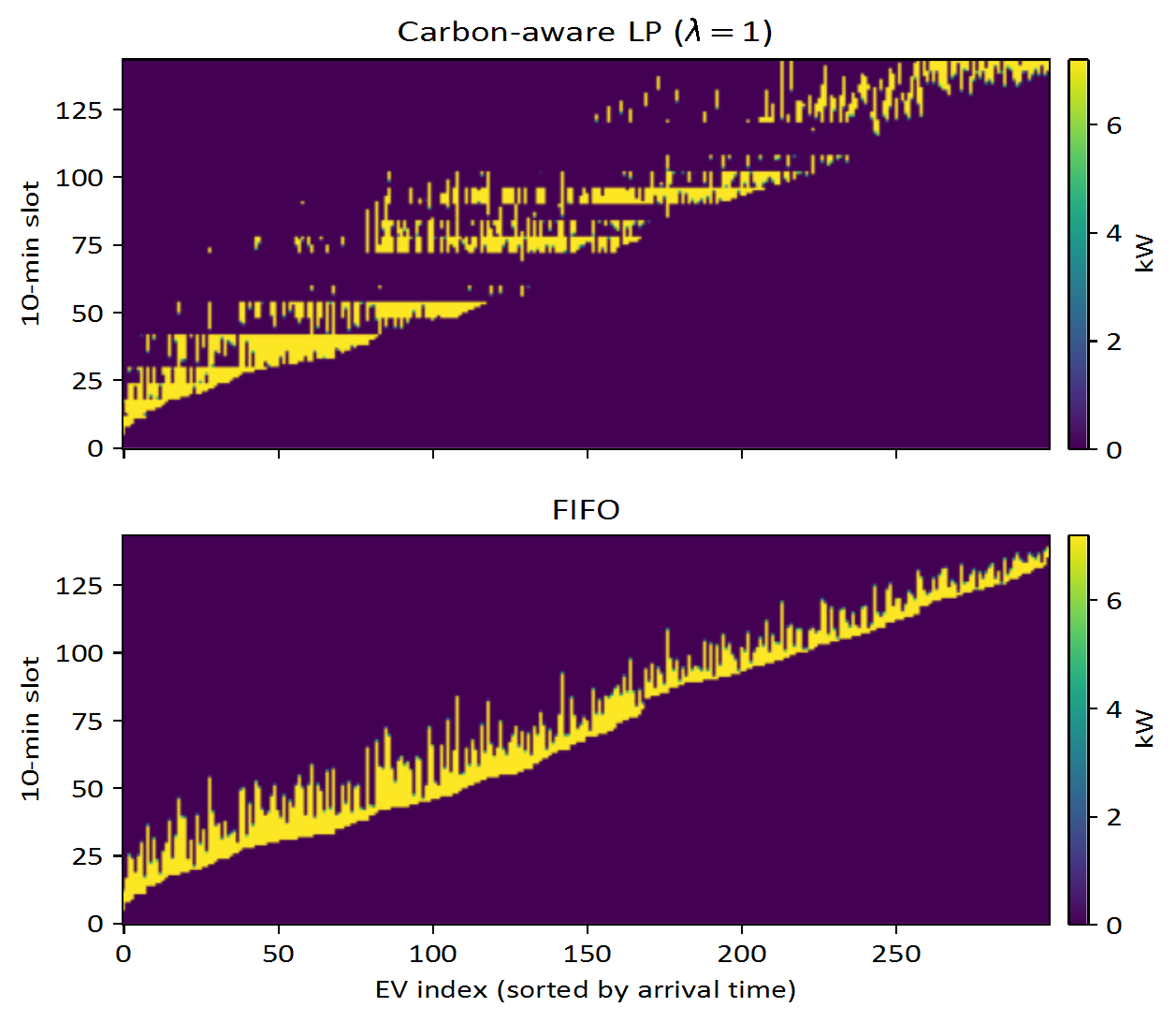}
    \caption{Charging heatmaps for the representative day. EVs are sorted by arrival time. The optimizer uses interruptions and resumptions, while FIFO follows a nearly monotone service pattern.}
    \label{fig:heatmaps}
\end{figure}

The aggregate power curves in Figure~\ref{fig:charge} hide an important structural difference between FIFO and the optimizer. Figure~\ref{fig:heatmaps} visualizes the slot-by-slot allocation matrix for the representative day. FIFO produces the expected staircase pattern: vehicles begin charging immediately after arrival and continue until their demand is satisfied. The carbon-aware LP instead interrupts and resumes many sessions, using user flexibility to move energy from carbon-intensive periods toward cleaner ones. This is exactly the type of temporal reshaping that a myopic arrival-ordered policy cannot reproduce.

\subsection{Pareto frontier and Monte Carlo study}
\begin{figure}[t]
    \centering
    \includegraphics[width=0.98\columnwidth]{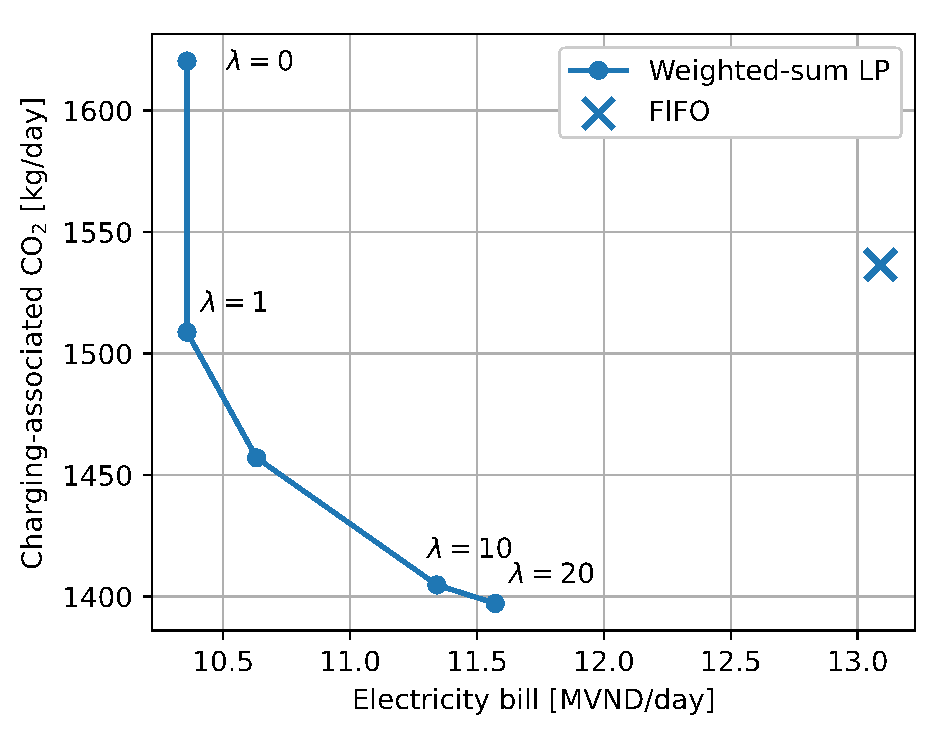}
    \caption{Supported Pareto frontier obtained by sweeping $\lambda$ in \eqref{eq:charge}. FIFO is shown for reference.}
    \label{fig:pareto}
\end{figure}

Figure~\ref{fig:pareto} shows the supported Pareto frontier for the representative day. Several nearby $\lambda$ values collapse to the same point, which is normal in weighted-sum scalarizations of LPs: the optimizer stays on one exposed extreme point until the scalar weight becomes large enough to activate a different supporting hyperplane. The key message is that FIFO is dominated by multiple Pareto points, while the cost-only LP is merely one frontier endpoint.

To verify that the conclusions are not tied to one arrival pattern, we simulate 100 random days. Each day serves on average 3.82~MWh of charging energy. Table~\ref{tab:mc} reports average electricity bill, charging-associated \COtwo{}, and mean solve time. The cost-only LP remains the cheapest policy, but also appears to be the ``dirtiest''. Setting $\lambda=1$ preserves the same 19.8\% bill reduction as the cost-only policy while decreasing emissions by 7.3\% relative to that cost-only schedule and by 2.5\% relative to FIFO. With $\lambda=10$, emissions drop by 9.3\% versus FIFO while the bill is still 12.6\% lower.

\begin{table}[t]
\caption{Average results over 100 random days\\(mean delivered energy: 3.82~MWh/day).}
\label{tab:mc}
\centering
\footnotesize
\begin{tabular}{lccc}
\toprule
Policy & Bill [MVND/day] & \COtwo{} [kg/day] & Solve time [s] \\
\midrule
FIFO & 13.03 & 1556.8 & -- \\
LP, $\lambda=0$ & 10.45 & 1637.3 & 0.030 \\
LP, $\lambda=1$ & 10.45 & 1518.2 & 0.033 \\
LP, $\lambda=10$ & 11.39 & 1412.4 & 0.034 \\
\bottomrule
\end{tabular}
\end{table}

The computational burden is very small: the dispatch LP in \eqref{eq:dispatch} solves in about 2~ms, and the charging LP in \eqref{eq:charge} solves in roughly 0.03~s. This is compatible with repeated day-ahead studies, sensitivity analyses, or online re-optimization when updated arrivals become available.

\subsection{Hydro sensitivity and seasonal interpretation}
Hydropower is the main low-carbon \emph{dispatchable} resource in the model.
We next perturb the hydro budget by scaling the initial storage, storage bounds, and hourly inflow by factors 0.8 and 1.2 around the nominal case, while keeping thermal and renewable capacities fixed. This is not a full seasonal study, but it is enough to reveal how hydrology reshapes the carbon signal seen by the charging station. Table~\ref{tab:hydro} reports the resulting mean system carbon intensity, coal share, and representative-day charging emissions.

\begin{table}[t]
\caption{Representative-day sensitivity to hydro availability.}
\label{tab:hydro}
\centering
\footnotesize
\setlength{\tabcolsep}{4pt}
\begin{tabular}{lccccc}
\toprule
Hydro case & Mean CI & Coal & FIFO & LP, $\lambda=1$ & LP, $\lambda=10$ \\
 & [g/kWh] & [\%] & [kg/day] & [kg/day] & [kg/day] \\
\midrule
0.8$\times$ hydro & 465.7 & 47.6 & 1737.0 & 1706.8 & 1580.4 \\
1.0$\times$ hydro & 413.0 & 41.0 & 1536.6 & 1508.8 & 1404.8 \\
1.2$\times$ hydro & 360.3 & 34.3 & 1346.9 & 1300.1 & 1248.2 \\
\bottomrule
\end{tabular}
\end{table}

The sensitivity highlights the distinct roles of renewables and hydro. PV mainly opens low-carbon windows around late morning and early afternoon, whereas hydro determines how much of the late-afternoon and evening demand can be served without additional coal. When hydro is scarce, the average carbon intensity rises from 413.0 to 465.7~g/kWh and coal share increases from 41.0\% to 47.6\%. In that harder case, the balanced scheduler still attains the same 10.36~MVND electricity bill as in the nominal scenario, while the more carbon-focused schedule remains 8.2\% cheaper and 9.0\% cleaner than FIFO.

When hydro is abundant, the grid becomes cleaner throughout the day and the price--carbon conflict weakens, but it does not disappear. The balanced policy yields 1300.1~kg/day, compared with 1346.9~kg/day for FIFO, and the $\lambda=10$ schedule reaches 1248.2~kg/day while still costing only 10.90~MVND/day, i.e., 16.7\% below FIFO. Hence the exact Pareto frontier is hydrology-dependent, as expected, yet the structural message is robust: updating the dispatch scenario is enough to adapt the charging signal, without changing the charging LP itself.

From an operator viewpoint, Table~\ref{tab:hydro} suggests a useful rule of thumb. Hydro-poor days mainly increase the penalty for charging into the evening peak, because coal and gas remain on the margin for longer; hydro-rich days, instead, create cleaner shoulder hours but do not eliminate the bill advantage of nighttime charging. The optimal response is therefore not a fixed heuristic such as ``always defer to the night'' or ``always absorb midday PV,'' but a schedule that explicitly tracks both the retail tariff and the supply mix.

This observation also clarifies why the \COtwo{} term is kept separate from the retail tariff instead of being absorbed into one ad hoc price signal. The electricity bill is driven by the station-facing TOU tariff, while the environmental signal is dispatch- and scenario-dependent. Keeping the two ingredients distinct makes the formulation easier to interpret and lets the same charging LP be reused under different hydrology, renewable availability, or carbon-price assumptions.

\subsection{Feedback sensitivity and limitations}
\begin{figure}[t]
    \centering
    \includegraphics[width=0.98\columnwidth]{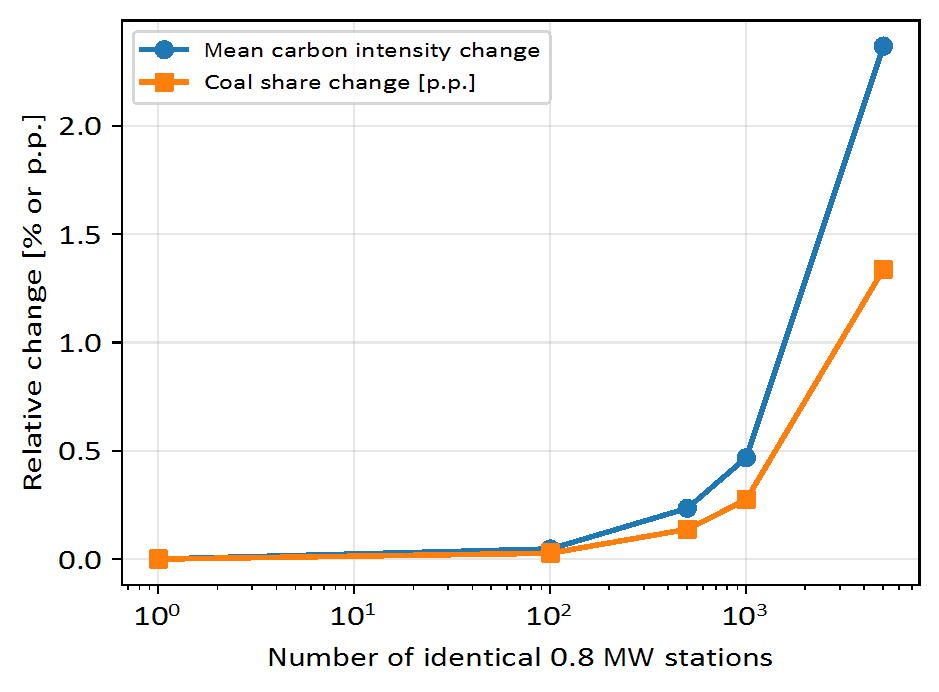}
    \caption{Ex-post feedback check: the optimized charging profile is replicated across many identical stations and added back to the dispatch layer. The one-way coupling is accurate for a single station and remains reasonable for moderate aggregations.}
    \label{fig:scale}
\end{figure}

The proposed decomposition should be interpreted carefully. Because the charging station is tiny relative to the national system in this study, using the dispatch only as an \emph{exogenous} carbon signal is defensible. To quantify this statement, Figure ~\ref{fig:scale} feeds the optimized hourly charging profile back into the dispatch layer after scaling it by the number of identical 0.8~MW stations. With one station, the change in mean daily carbon intensity is essentially zero. Even with 1000 identical stations, the mean carbon intensity rises by only 0.47\% and coal share by 0.28 percentage points. The decoupled approximation becomes visibly less reliable only for extremely large aggregations (5000 identical stations in this stylized calculation, roughly 4~GW of coincident feeder capacity).

A second limitation is perfect foresight: the scheduler assumes the arrival/departure windows and the day-ahead carbon signal are known. The present results therefore represent an optimistic upper bound, albeit a useful one. Extending problem \eqref{eq:charge} with stochastic or robust constraints is straightforward and aligns with recent robust EV-charging formulations \cite{Sun2020,Calafiore2025}.

\section{Conclusions}
This paper presented a tractable framework for carbon-aware EV smart charging that couples a simplified day-ahead power-system dispatch model with an optimization-based charging scheduler. 
The study is grounded on realistic regional data from Vietnam.
The dispatch layer provides an hourly estimate of the carbon intensity of electricity supply by accounting for the interplay among thermal generation, renewable availability, hydropower, imports, and demand, while the charging layer exploits this signal jointly with time-of-use electricity prices to schedule aggregate EV charging.

The numerical results show that the proposed approach can substantially reshape charging demand away from carbon-intensive hours while preserving economic efficiency and operational feasibility. Compared with a FIFO rule, the optimization-based policies achieve lower charging cost and lower emissions, and the weighted formulation makes the cost--emissions trade-off explicit through a tunable parameter. The additional sensitivity and Monte Carlo analyses further indicate that the main qualitative conclusions are robust to variations in hydropower availability and fleet-level charging uncertainty.

Overall, the results support the view that even a lightweight dispatch-informed signal can materially improve charging decisions in practice, without requiring a full market simulator or plant-level commitment model. Future work will consider finer network and reserve constraints, stochastic renewable and inflow models, and larger heterogeneous EV fleets with distributed implementation and real-time recourse.

\balance

\end{document}